\documentclass{article}

\usepackage[final,nonatbib]{neurips_2018}

\usepackage[utf8]{inputenc} 
\usepackage[T1]{fontenc}    
\usepackage{hyperref}       
\usepackage{url}            
\usepackage{booktabs}       
\usepackage{amsfonts}       
\usepackage{nicefrac}       
\usepackage{microtype}      
\usepackage{bm}
\usepackage{graphicx}

\title{Quantifying Visual Image Quality: A Bayesian View}

\author{%
  Zhengfang~Duanmu \\
  University of Waterloo\\
  Waterloo, ON, N2L 3G1 \\
  \texttt{zduanmu@uwaterloo.ca} \\
   \And
  Wentao~Liu \\
  University of Waterloo\\
  Waterloo, ON, N2L 3G1 \\
  \texttt{w238liu@uwaterloo.ca} \\
   \AND
  Zhongling~Wang \\
  University of Waterloo\\
  Waterloo, ON, N2L 3G1 \\
  \texttt{zhongling.wang@uwaterloo.ca} \\
   \And
  Zhou~Wang \\
  University of Waterloo\\
  Waterloo, ON, N2L 3G1 \\
  \texttt{zhou.wang@uwaterloo.ca} \\
}

\begin{document}

\maketitle

\begin{abstract}
Image quality assessment (IQA) models aim to establish a quantitative relationship between visual images and their perceptual quality by human observers.
IQA modeling plays a special bridging role between vision science and engineering practice, both as a test-bed for vision theories and computational biovision models, and as a powerful tool that could potentially make profound impact on a broad range of image processing, computer vision, and computer graphics applications, for design, optimization, and evaluation purposes.
IQA research has enjoyed an accelerated growth in the past two decades.
Here we present an overview of IQA methods from a Bayesian perspective, with the goals of unifying a wide spectrum of IQA approaches under a common framework and providing useful references to fundamental concepts accessible to vision scientists and image processing practitioners.
We discuss the implications of the successes and limitations of modern IQA methods for biological vision and the prospect for vision science to inform the design of future artificial vision systems\footnote{The detailed model taxonomy can be found at \url{http://ivc.uwaterloo.ca/research/bayesianIQA/}.}.
\end{abstract}

\section{Introduction}
The goal of research in objective image quality assessment (IQA) is to develop computational models that can automatically predict perceived image quality by human observers.
Although assessing image quality appears to be an easy task for humans, the underlying mechanisms are not well understood, making model prediction a challenging task.
The research in IQA plays a special role as a bridge between vision science and engineering practice.
On the one hand, IQA offers an excellent test-bed for evaluating vision theories and computational biovision models.
In contrast to many traditional vision research that typically focuses on qualitative explanations of certain observed vision behaviors, the task of IQA provides a strong test for the ``quantitative'' prediction power of visual processing hypotheses with a broad space of interests.
On the other hand, IQA is an essential component in all image processing, computer vision, and computer graphics applications for which human eyes are the ultimate receivers.
IQA models are not only used as the criteria to evaluate and compare algorithms and systems, but also serve as the guide to drive the design and optimization of perceptually inspired algorithms and systems.
Therefore, advancement in IQA research may make fundamental impact on the development of numerous real-world technologies that involve image processing, computer vision, and computer graphics.

There has been an accelerated development in IQA research, especially in the past 20 years.
A good number of subject-rated image quality databases have been constructed and made public that enable IQA algorithms to be trained and tested for a variety of application scenarios~\cite{athar:19}.
Several design principles have emerged and have been shown to be effective at creating IQA algorithms, many of which are well correlated with perceptual image quality when tested using the current public image quality databases~\cite{athar:19}.
The achievement is worth celebrating, especially when compared with what we had 20 years ago, when simple numerical measures such as the peak-signal-to-noise-ratio (PSNR), a direct mapping of the mean-squared-error (MSE) to the logarithm scale, could compete on a par with then state-of-the-art perceptual quality metrics~\cite{vqeg:2000}.

Despite the demonstrated success, several outstanding challenges remain in the fundamentals of IQA research.
First, a well-structured problem formulation is missing that not only provides a unified framework to understand the connections between IQA models, but also identifies potential ways for future development.
Second, the multi-discipline nature of IQA research gives rise to misconceptions and ambiguities concerning some basic IQA terminologies.
In particular, visual quality is frequently confused with perceptual similarity, perceptual metric, and image aesthetics, resulting in vague optimization goals, inconsistent psychophysical experimental protocols, and inadequate evaluation criteria.
Third, many algorithms are derived in ad-hoc manner where assumptions are implicit, making it extremely challenging to fairly evaluate competing hypotheses and recognize their limitations.
Fourth, while it seems obvious that a successful IQA model has to relate to the visual processing system in some way, many methods fail to draw a connection to vision science.
As a result, it is often difficult to make an intuitive sense of how and why an IQA model works.
With a growing number of new IQA models emerging each year, we have seen more ``symptoms'' arising from the aforementioned fundamental issues.
For example, some recent IQA techniques are reported as ``unreasonably effective'' and ``unexpectedly powerful''~\cite{zhang:18}.


The Bayesian theory has found profound applications in vision science by offering a principled yet simple computational framework for perception that accounts for a large number of perceptual effects and visual behaviors~\cite{knill:96}.
Meanwhile, Bayesian inference and estimation theories have been employed extensively in a wide variety of computer vision, image processing, computer graphics, and machine learning methods~\cite{prince:12}.
In this paper, we attempt to bridge the gap between the two, by laying out a generic conceptual framework for quantifying image quality from a Bayesian perspective.
We provide a general formulation of the objective IQA problem, highlighting a branch of statistical models that underpin the existing IQA methods.
We discuss two types of Bayesian networks for IQA with distinct definitions on visual image quality.
We also identify common source of prior information for developing artificial vision systems, and discuss a series of examples in which researchers have used a specific type of prior knowledge.
Finally, we describe existing evaluation criteria, from intuitive sanity check to sophisticated analysis-by-synthesize approaches.
Given the space constraints, we do not dive into great technical details, but point interested readers to further readings~\cite{athar:19,chandler:13,wang:06a,wang:11,wang:16a,zhai:20}.

\section{Bayesian View of Image Quality Assessment}
The goal of IQA is to determine the subjective quality rating $y$ given an image ${\bf x}$.
The problem can be formulated as a Bayesian inference problem, where the objective is to determine the probability distribution $p(y|{\bf x})$, which may be followed by a decision making process that generates a deterministic estimate of $y$.
There are generally two distinct approaches to solving the inference problem.

The first approach firstly solves the inference problem by determining the quality level-conditional densities $p({\bf x}|y)$ for each quality level $y$ and the prior label probabilities $p(y)$.
Then one can use Bayes' theorem in the form
\begin{equation}
    p(y | {\bf x}) = \frac{p({\bf x}|y) p(y)}{p(\bf x)},
\end{equation}
to find the posterior quality distribution $p(y | {\bf x})$~\cite{xue:13a}. The denominator in Bayes’ theorem can be found in terms of the quantities appearing in the numerator, because
\begin{equation}
    p({\bf x}) = \int p({\bf x}|y) p(y) dy.
\end{equation}
The models generated from this approach is known as \textit{generative models}, because by sampling from them it is possible to generate synthetic data points in the input space.
However, due to the lack of training data and effective learning methods, generative models have not drawn much attention from IQA researchers.
As a result, we focus on the second approach in this review.

Alternatively, the second approach aims to determine the posterior quality probabilities $p(y|{\bf x})$ directly.
This approach is simpler in the sense that we do not need to model the image space, of which we only have limited understanding.
However, building an accurate model of $p(y|{\bf x})$ still requires sampling and performing subjective tests on all possible images, neither of which is feasible in practice.
Therefore, most existing IQA models are focused on the following problem:
Given a set of training data $\mathcal{D}$ comprising $n$ input images (and optionally some side-information) ${\bf X} = ({\bf x}_1, . . . , {\bf x}_n)$ and their corresponding target quality scores ${\bf y} = (y_1, . . . , y_n)$, find a posterior quality distribution $p(y|\mathbf{x}, \mathcal{D})$ that best approximates $p(y|\mathbf{x})$ in the human visual system (HVS).
It should be noted that $p(y|\mathbf{x}, \mathcal{D})$ can be regarded as a point estimate of $p(y|\mathbf{x})$ as the latter would be fully recovered by $\int p(y|\mathbf{x}, \mathcal{D})p(\mathcal{D})d\mathcal{D}$ if we sample all possible data $\mathcal{D}$.
The problem is further simplified by assuming the training data are independent and identically distributed, so that the predictive distribution can be parametrized~\cite{de:17} as
\begin{equation}\label{eq:prediction}
    p(y|{\bf x}, \mathcal{D}) = \int p(y|{\bf x}, {\bm{\theta}}) p(\bm{\theta}|\mathcal{D}) d\bm{\theta},
\end{equation}
where $\bm{\theta}$, $p(y|{\bf x}, {\bm{\theta}})$ and $p(\bm{\theta}|\mathcal{D})$ represent the parameters of the HVS model, the quality rating generation process and the posterior distribution over parameters, respectively.
Given the enormous space of $\bm{\theta}$, the computation of the integral in Equation~\ref{eq:prediction} is prohibitively expensive.
As a result, a common practice is to approximate the predictive distribution $p(y|{\bf x}, \mathcal{D})$ by a point estimate $p(y|{\bf x}, \bm{\theta}^*)$, where
\begin{equation}\label{eq:map_theta}
    \bm{\theta}^* = \arg\max_{\bm{\theta}} p(\bm{\theta}|\mathcal{D}) = \arg\max_{\bm{\theta}} p({\bf y}|{\bf X}, \bm{\theta}) p(\bm{\theta}).
\end{equation}
The specific form of the likelihood function $p({\bf y}|{\bf X}, \bm{\theta})$ is not known in practice.
To fully specify the problem, it is usually assumed that the likelihood function follows a Gaussian distribution
\begin{equation}\label{eq:gauss_assump}
    p(y | {\bf x}, \bm{\theta}, \beta) = \mathcal{N}(y | f({\bf x}; \bm{\theta}), \beta),
\end{equation}
where $f({\bf x}; \bm{\theta})$ and $\beta$ represent the mean and variance of the Gaussian distribution, respectively.
It is easy to show that the maximum likelihood solution of $\bm{\theta}$ is equivalent to the best least-square solution with respect to the mean opinion score (MOS) under this assumption.

Direct estimation of $\bm{\theta}$~\cite{kang:14} from a set of training data is problematic, because of the fundamental conflict between the enormous size of the image space and the limited scale of affordable subjective testing.
Specifically, a typical ``large-scale'' subjective test allows for a maximum of several hundreds or a few thousands of test images to be rated.
Given the combination of source images, distortion types and distortion levels, realistically only a few dozens of source images (if not fewer) can be included, which is the case in all known subject-rated databases.
By contrast, digital images live in an extremely high dimensional space, where the dimension equals the number of pixels, which is typically in the order of hundreds of thousands or millions.
Therefore, a few thousands of samples that can be evaluated in a typical subjective test are deemed to be extremely sparsely distributed in the space.
Furthermore, it is difficult to justify how a few dozens of source images can provide a sufficient representation of the variations of real-world image content.
As a result, the fundamental problem in the objective IQA is to develop a meaningful prior parameter distribution $p(\bm{\theta})$, which encodes the configuration of the HVS.

\begin{figure}[t!]
\includegraphics[width=\textwidth]{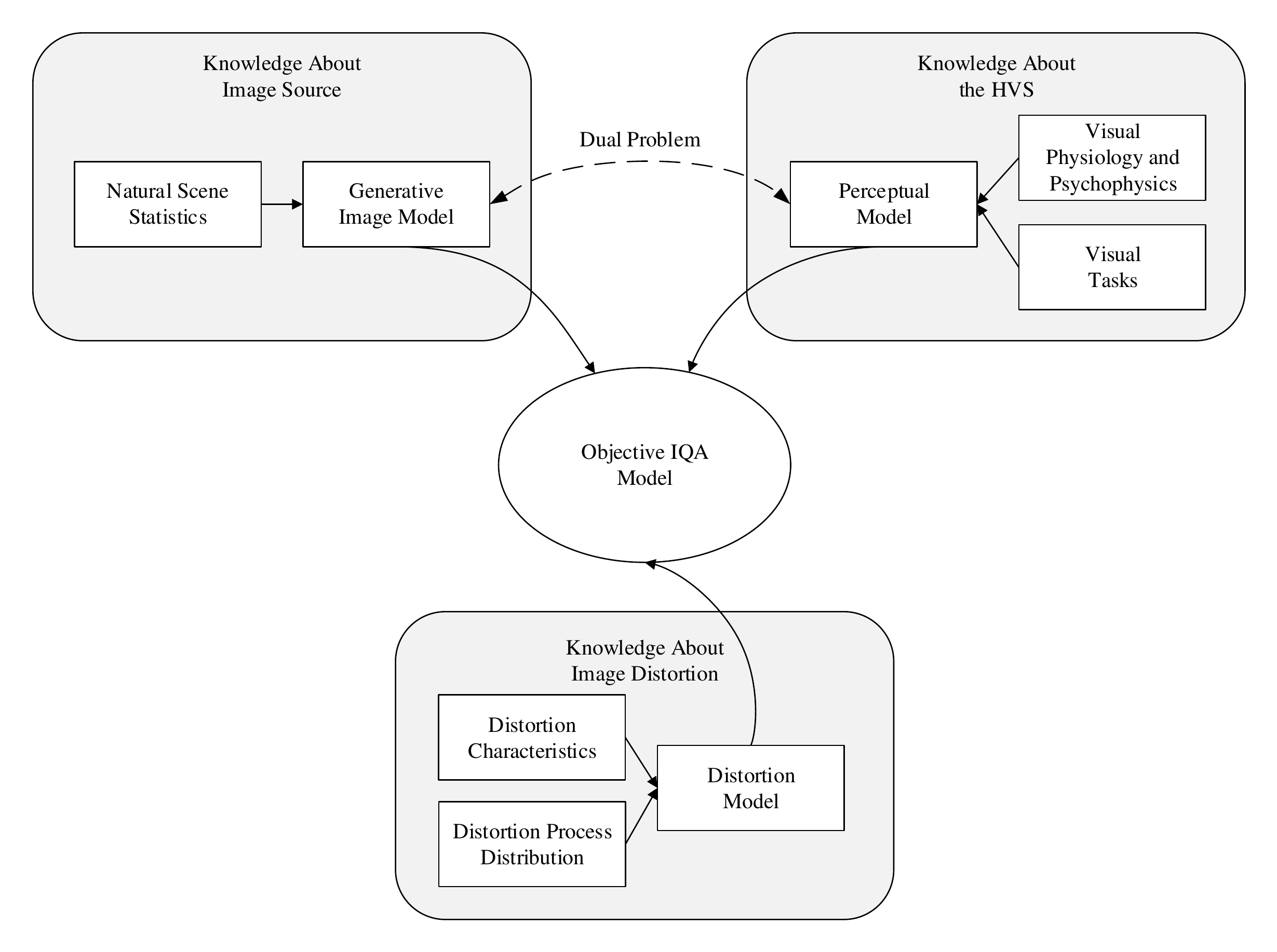}
\caption{Knowledge map of objective IQA.}
\label{fig:knowledge}
\end{figure}

Over the past decades, various IQA models have been developed where the key difference lies in the assumptions about the prior distribution $p(\bm{\theta})$.
In general, three types of knowledge may be used for the design of image quality measures, as shown in Figure~\ref{fig:knowledge}.
Most systems attempt to incorporate knowledge about the HVS, which can be further divided into bottom-up knowledge and top-down assumptions.
The former includes the computational models that have been developed to account for a large variety of physiological and psychophysical visual experiments~\cite{heeger:92,olshausen:97}.
The latter refers to those general hypotheses about the overall functionalities of the HVS~\cite{wang:04a}.

Knowledge about the possible distortion processes is another important information source in the design of objective IQA models.
This type of information generally includes the appearance of certain distortion pattern and the distribution of distortion processes in practice.
For example, one can explicitly construct features that are aware of particular artifacts, such as blocking~\cite{wang:02a}, blurring~\cite{wang:04b}, and ringing~\cite{marziliano:04}, and then assign penalties to these distortions.
Also, it is much easier to create distorted image examples that can be used to train these models, so that more accurate image quality prediction can be achieved.
This type of knowledge is typically deployed in IQA models that are specifically designed to handle a specific artifact type.

The third type is knowledge about the visual world to which we are exposed.
It essentially summarizes what natural images should, or should not, look like.
It is known that there exist strong statistical regularities of the natural images~\cite{simoncelli:01}.
If an observed image significantly violates such statistical regularities, then the image is considered unnatural and is presumably of low quality.
The statistical properties of natural images, which are often referred to as Natural Scene Statistics (NSS), have profound impact on the research in the general-purpose IQA~\cite{wang:11} and are still making significant impacts in the deep learning era.
In computational neuroscience, it has long been conjectured that the HVS is highly adapted to the natural visual environment~\cite{barlow:61}, and therefore, the modeling of natural scenes and the HVS are dual problems~\cite{sheikh:05}.

\section{Full-Reference Image Quality Assessment}
\begin{figure}[t!]
\includegraphics[width=\textwidth]{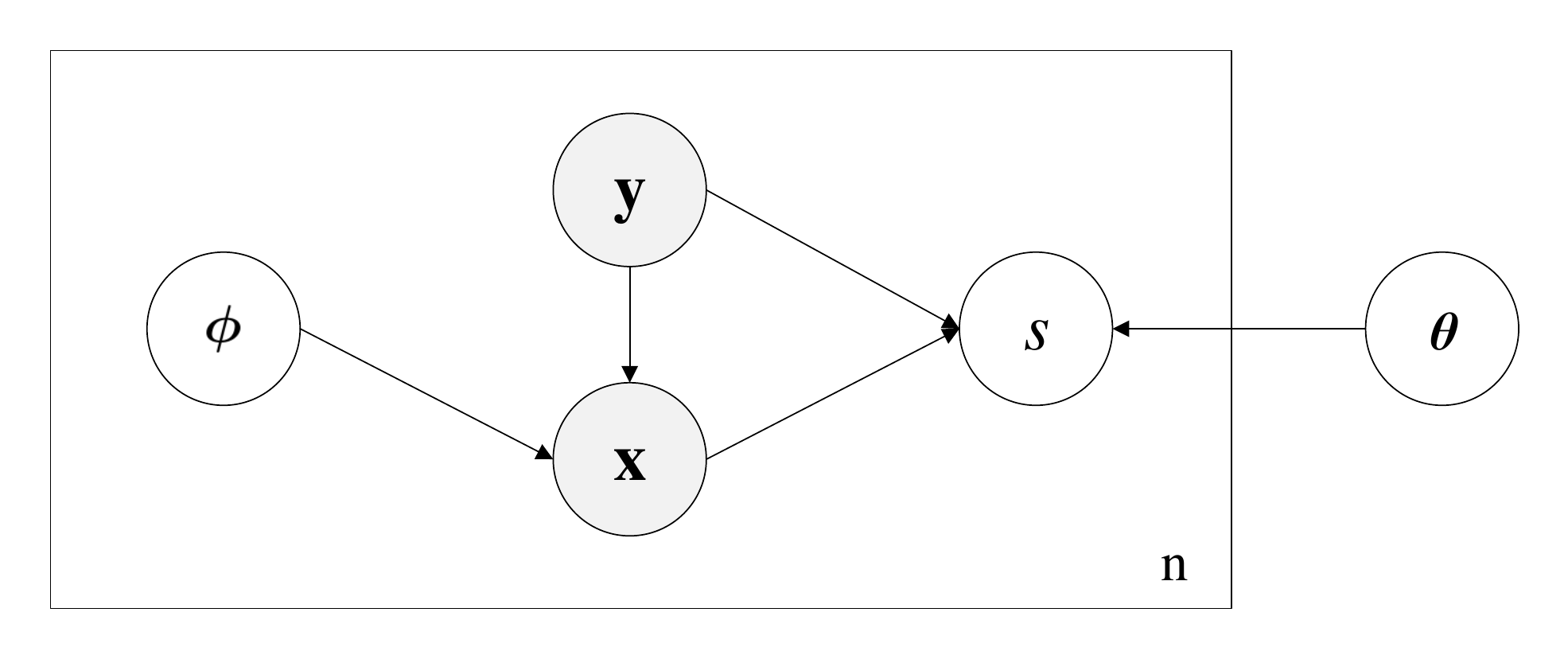}
\caption{Graphical model representation of FR IQA models. The box is ``plate'' representing replicates. Each node represents a random variable (or group of random variables), and the links express probabilistic relationships between these variables. The observable variables are shaded in color.}
\label{fig:friqa}
\end{figure}
Pioneering work on perceptual image processing and IQA dates back at least to the 1970's, when Mannos and Sakrison investigated a family of visual fidelity measures in the context of rate-distortion optimization~\cite{mannos:74}.
Since then, researchers started to connect image quality with perceptual fidelity.
Assuming the test image is generated from a pristine image, early IQA methods assess image quality by comparing the two images and producing a quantitative score that describes the degree of similarity/fidelity or, conversely, the level of error/distortion between them.
The equivalence between image quality and perceptual fidelity makes intuitive sense, because the test image is more likely to have high quality as it looks closer to the reference image.
Although ``image quality'' is frequently used for historical reasons, the more precise term for this type of metric would be image similarity or fidelity measurement, or full-reference (FR) IQA.

The FR IQA problem can be explained by Equation~\ref{eq:prediction}, where each observation ${\bf x}$ consists of a pair of images.
Given an original image of acceptable (or perhaps pristine) quality ${\bf x}_r$ and its altered version, a test image ${\bf x}_t$, that undergoes a distortion process $g(\cdot; \boldsymbol{\phi})$, FR IQA models aim to estimate the quality conditional probability distribution $p(y|{\bf x}_t, {\bf x}_r, \bm{\theta})$.
The probabilistic graphical model of FR IQA models is shown in Figure~\ref{fig:friqa}.
By assuming the quality label generation process follows a Gaussian distribution
\begin{equation}\label{eq:gauss_assump2}
    p(y | {\bf x}_t, {\bf x}_r, \bm{\theta}, \beta) = \mathcal{N}(y | d({\bf x}_t, {\bf x}_r; \bm{\theta}), \beta)
\end{equation}
and a point estimate of $\bm{\theta}$, we reduce the FR IQA problem to finding a deterministic perceptual similarity measure $d({\bf x}_t, {\bf x}_r; \bm{\theta})$, where we have encoded our prior knowledge by $\bm{\theta}$.

\begin{figure}[t!]
\includegraphics[width=0.6\paperwidth]{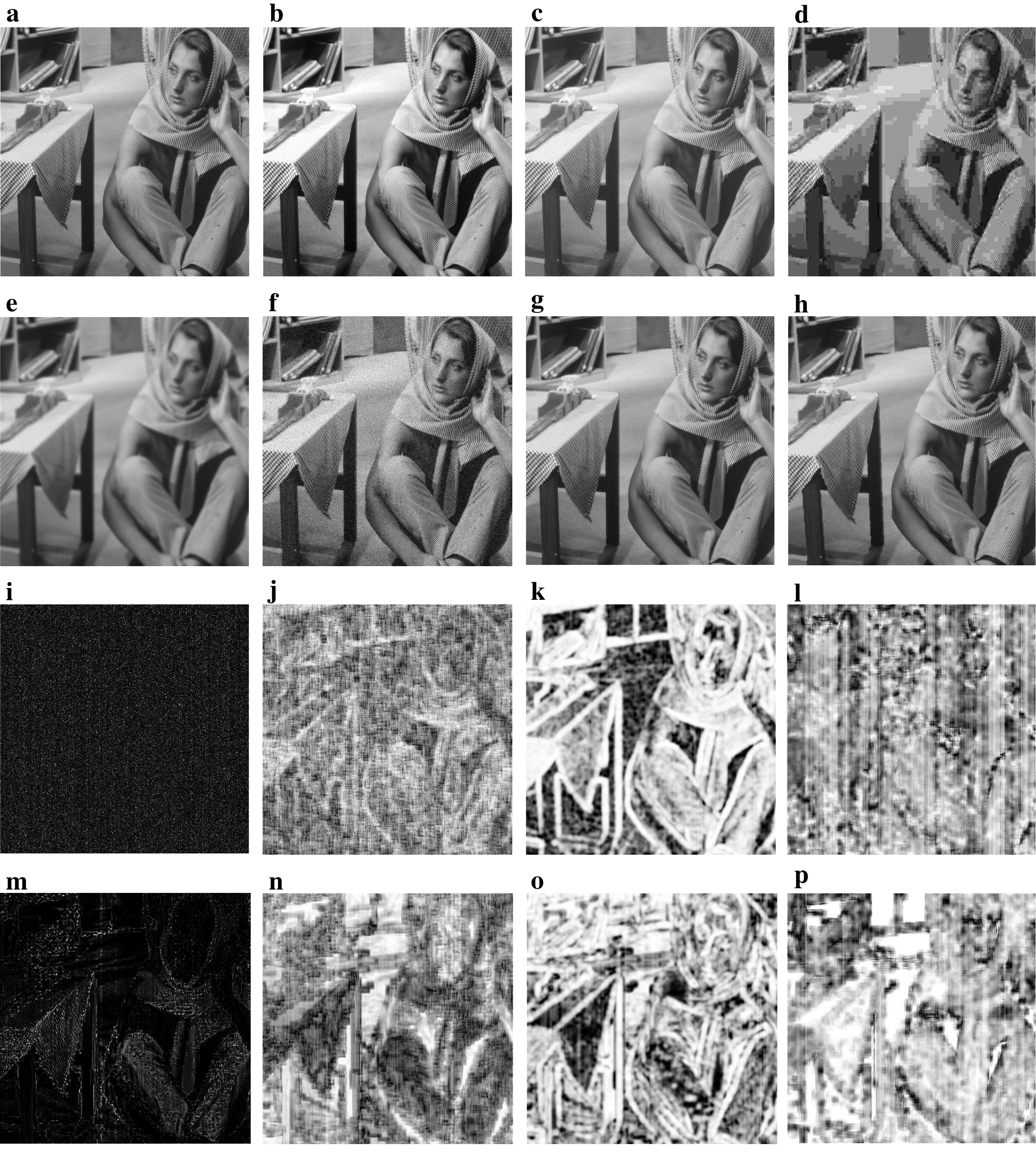}
\caption{(a) The original ''Barbara'' image. (b)-(h) Comparison of ``Barbara'' images with different types of distortions, all with $\mathrm{MSE} = 300$. (i)-(p) Quality maps of (f) and (d) generated from different FR IQA algorithms. (b) Contrast-stretched image, $\mathrm{SSIM}=0.966$, $\mathrm{VIF}=1.115$, $\mathrm{NLPD}=0.142$. (c) Mean-shifted image, $\mathrm{SSIM}=0.982$, $\mathrm{VIF}=1$, $\mathrm{NLPD}=0.020$. (d) JPEG compressed image, $\mathrm{SSIM}=0.740$, $\mathrm{VIF}=0.153$, $\mathrm{NLPD}=0.427$. (e) Blurred image, $\mathrm{SSIM}=0.792$, $\mathrm{VIF}=0.247$, $\mathrm{NLPD}=0.306$. (f) White Gaussian noise contaminated image, $\mathrm{SSIM}=0.803$, $\mathrm{VIF}=0.342$, $\mathrm{NLPD}=0.364$. (g) Vertical translated image, $\mathrm{SSIM}=0.637$, $\mathrm{VIF}=0.096$, $\mathrm{NLPD}=0.667$. (h) Rotated image, $\mathrm{SSIM}=0.427$, $\mathrm{VIF}=0.062$, $\mathrm{NLPD}=0.943$. (i) MSE map of (f). (j) NLPD map of (f). (k) SSIM map of (f). (l) VIF map of (f). (m) MSE map of (d). (n) NLPD map of (d). (o) SSIM map of (d). (p) VIF map of (d).}
\label{fig:example}
\end{figure}
The simplest and widely used FR IQA is the MSE, which still remains a popular quantitative criterion for assessing image quality~\cite{wang:09}.
Suppose that ${\bf x}_t=\{x_{t,i}|i=1, 2, ..., m\}$ and ${\bf x}_r=\{x_{r,i}|i=1, 2, ..., m\}$ are distorted and reference images, where $m$ is the number of pixels and $x_{t,i}$ and $x_{r,i}$ are the values of the $i$-th samples in ${\bf x}_t$ and ${\bf x}_r$, respectively.
The MSE between the images is
\begin{equation}\label{eq:mse}
    d_{\mathrm{MSE}} = \frac{1}{m} \sum_{i=1}^{m} (x_{t,i} - x_{r,i})^2.
\end{equation}
In this case, the prior knowledge is encoded by the functional form of MSE, which can be denoted by $\bm{\theta}_{\mathrm{MSE}}$.
Since the functional form is deterministic, we have $p(\bm{\theta}=\bm{\theta}_{\mathrm{MSE}})=1$ and $p(\bm{\theta}=\bm{\theta}')=0$ for any function $\bm{\theta}' \neq \bm{\theta}_{\mathrm{MSE}}$.
Consequently, the posterior distribution $p(\bm{\theta}|\mathcal{D})$ converges to the prior distribution $p(\bm{\theta})$ for any likelihood function and dataset as long as $p(\mathcal{D}|\bm{\theta}_{\mathrm{MSE}}) > 0$.
The use of MSE as an image quality measure is appealing because it is simple to calculate, has clear physical meanings, and is mathematically convenient in the context of optimization.
Unfortunately, MSE is not very well matched to perceived visual quality~\cite{wang:09}.
An illustrative example is shown in Figure~\ref{fig:example}a-h, where the original ``Barbara'' image is altered with different distortions, each adjusted to yield nearly identical MSE relative to the original image.
Despite this, the images can be seen to have drastically different perceptual quality.
The failure of MSE in predicting image quality arises from neglecting the knowledge about natural images, distortion processes, and the HVS.
In the last four decades, a great deal of effort has gone into the development of FR IQA methods that take advantage of these knowledge.
We summarize these techniques in the subsequent section.

\subsection{Error Visibility Paradigm}
Given the reference image, it is straightforward to compute the numerical errors between the reference and test images.
Error visibility methods predict image quality as the visibility of such errors based on psychophysical and physiological models of the HVS.
Almost all early well-known perceptual image quality models~\cite{carlson:80,daly:92,lubin:93,lubin:95,mannos:74,safranek:89,teo:94,watson:87,watson:93,watson:97} followed this error visibility paradigm, which was well laid out as early as 1993~\cite{ahumada:93} and later refined~\cite{wang:04a}.
Specifically, it has been found that the HVS is relatively insensitive to certain types of visual patterns.
First of all, the HVS is known to have different sensitivity to the spatial frequency content in visual stimuli.
The relationship between the sensitivity of the HVS and the spatial frequency content in visual stimuli can be modeled by the contrast sensitivity function (CSF)~\cite{watson:93}, which peaks at a spatial frequency around four cycles per degree of visual angel and drops significantly with both increasing and decreasing frequencies.
For example, it can be observed that the crossing pattern on the bamboo chair looks clearer than the high frequency texture on the scarf in Figure~\ref{fig:example}a.
Second, the presence of one signal can sometimes reduce the visibility of another image component.
As an illustrative example, the noise signal on the scarf and tablecloth appears to be less visible than the distortion on the girl's face in Figure~\ref{fig:example}f, although the Gaussian noise is applied uniformly across the image.
The phenomenon is known as the contrast masking effect.
In general, a masking effect is strongest when the signal and the masker have similar spatial location, frequency content, and orientations as evident by Figure~\ref{fig:example}b.
Third, the perception of luminance obeys Weber's law, which can be expressed mathematically as $\frac{\Delta L}{L} = C$, where $L$ is the background luminance, $\Delta L$ is the just noticeable incremental luminance over the background by the HVS, and $C$ is a constant called the Weber fraction.
The effect can be observed in Figure~\ref{fig:example}f, where the noise on the leg of the table appears to be more noticeable than the noise on the floor.
Motivated by the different sensitivity of the HVS to visual stimuli, a large number of IQA models in the literature share a similar error visibility paradigm, although they differ in detail.
\begin{figure}[t!]
\includegraphics[width=0.7\paperwidth]{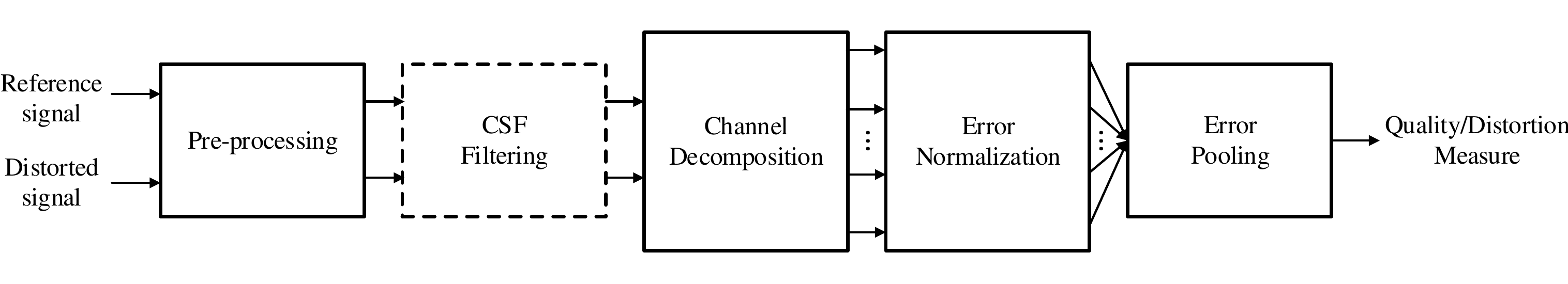}
\caption{A prototypical quality assessment system based on error sensitivity. CSF: Contrast sensitivity function. Image by courtesy of Wang \textit{et al.}~\cite{wang:04a}.}
\label{fig:error_visibility}
\end{figure}
Figure~\ref{fig:error_visibility} shows a generic error visibility IQA system framework.
The stages of the diagram are as follows.
\begin{itemize}
    \item Pre-processing: This stage typically performs a variety of basic operations to transform input images into the desired format, including spatial registration, color space transformation, point-wise non-linearity, and point spread function (PSF) filtering that mimics eye optics.
    \item CSF Filtering: Some FR IQA models weight the image component according to the CSF immediately after the pre-processing stage (typically implemented using a linear filter that approximates the frequency response of the CSF), while other error visibility models implement CSF as a base-sensitivity normalization factor after channel decomposition.
    \item Channel Decomposition: A large number of neurons in the primary visual cortex are tuned to visual stimuli with specific spatial locations, frequencies, and orientations.
    Motivated by the observation, these IQA methods have been using localized, band-pass, and oriented linear filters to decompose the input images into multiple channels.
    A number of signal decomposition methods have been used for IQA, including Fourier decomposition~\cite{mannos:74}, Gabor decomposition~\cite{nielsen:85,taylor:97}, local block-DCT transform~\cite{watson:93}, quadrature mirror filter bank~\cite{safranek:89}, separable wavelet transform~\cite{bradley:99,chandler:07,lai:00,teo:94,watson:97}, polar separable wavelet transform~\cite{watson:87}, and hexagonal orthogonal-oriented pyramid~\cite{watson:89}.
    \item Error Normalization: The error between the decomposed signals in each channel may be normalized by the CSF, and may also be normalized according to a certain masking model, which takes into account the effects of luminance masking and contrast masking.
    The normalization mechanism may be implemented as a spatially adaptive divisive normalization process~\cite{heeger:92}, and may also be implemented as a spatially varying thresholding function in a channel to convert the error into units of just noticeable difference.
    The visibility threshold at each point is calculated based on the energy of the reference and/or distorted coefficients in a neighborhood (which may include coefficients from within a spatial neighborhood~\cite{laparra:16} of the same channel as well as other channels~\cite{daly:92}) and the base-sensitivity for that channel.
    \item Error Pooling: The final stage of FR IQA models combine the normalized error signals over the spatial extent of the image, and across different channels, into a single scalar measure, which describes the overall quality of the distorted image.
    Most error pooling takes the form of a Minkowski norm as follows~\cite{ahumada:93,wang:06c}:
    \begin{equation}\label{eq:minkowski}
    E = \big(\sum_u \sum_v  |e_{u, v}|^{\gamma} \big)^{1/\gamma},
    \end{equation}
    where $e_{u, v}$ is the normalized error of the $u$-th coefficient in the $v$-th channel and $\gamma$ is a constant exponent chosen empirically.
\end{itemize}

Figure~\ref{fig:example} shows the quality scores of the ``Barbara'' image set and the quality map of the white Gaussian noise contaminated image generated by a state-of-the-art error visibility-based IQA model named Normalized Laplacian Pyramid Distance (NLPD)~\cite{laparra:16}, whose error normalization module is learned from subjective labeled data.
The predicted quality has a much higher correlation with human perception than MSE.

\subsection{Structural Similarity Paradigm}
The error visibility paradigm has received broad acceptance in real-world image processing applications.
However, it is important to realize the limitations of these methods.
A summary of some of the potential problems is as follows.
\begin{itemize}
    \item Most error visibility IQA models are based on linear or quasi-linear operators that have been characterized using restricted and simplistic stimuli such as spots, bars, or sinusoidal gratings.
    This is problematic for two reasons.
    First, the HVS consists of many non-linear units that is too complex to model precisely.
    Second, the stimuli used in the psychophysical experiments are much simpler than natural images, which can be thought of as a superposition of a large number of simple patterns.
    As a result, the generalization capability of these models remains limited.
    \item Not every error signal leads to quality degradation.
    Contrast enhancement gives an obvious example (Figure~\ref{fig:example}b), in which the difference between an original image and a contrast-enhanced image may be easily discerned, but the perceptual quality is not degraded.
    \item The error normalization module in error visibility models relies on psychophysical experiments that are specifically designed to estimate the just noticeable difference.
    However, there has been little evidence whether such near-threshold models can be generalized to characterize perceptual distortions significantly larger than threshold levels, as is the case in a majority of image processing situations.
    \item The Minkowski-based error pooling implicitly assumes that errors at different locations are statistically independent.
    However, such dependency cannot always be completely eliminated by linear channel decomposition and masking models.
\end{itemize}

To overcome these challenges, a different approach was taken by making use of the knowledge about the overall functionality of the HVS~\cite{wang:02b,wang:04a}.
The major assumption behind the structural similarity paradigm is that the HVS is highly adapted to extract structural information from the viewing field.
It follows that a measurement of structural similarity (or distortion) should provide a good approximation to perceptual image quality.
To convert the structure similarity paradigm into an IQA algorithm, it is necessary to define what structural/nonstructural distortions are and how to separate them.

Pioneering the structural similarity approach, Wang~\textit{et al.} proposed to define the nonstructural distortions as those distortions that do not modify the structure of objects in the visual scene, and all other distortions to be structural distortions~\cite{wang:02b}.
Figure~\ref{fig:example} is instructive in this regard.
Although the contrast enhanced/mean shifted distorted images can be easily distinguished from the reference image, the distorted images preserve virtually all of the essential information composing the structures of the objects in the image.
Indeed, the reference image can be recovered perfectly via a simple point-wise affine transformation.
As a result, luminance shift and contrast change are considered as nonstructural distortions, independent of other structural distortions.

This motivated a spatial domain implementation of the structural similarity idea called the Structural SIMilarity (SSIM) index~\cite{wang:04a}.
The system separates the task of similarity measurement into three independent comparisons: luminance, contrast and structure.
First, the local luminance of distorted and reference images are estimated by the mean intensity $\mu_{x_t}$ and $\mu_{x_r}$.
The luminance similarity between the two images is defined as
\begin{equation}\label{eq:ssim_l}
    l(\textbf{x}_t, \textbf{x}_r) = \frac{2\mu_{x_t}\mu_{x_r} + C_1}{\mu_{x_t}^2 + \mu_{x_r}^2 + C_1},
\end{equation}
where the constant $C_1$ is included to avoid instability when $\mu_{x_t}^2 + \mu_{x_r}^2$ is very close to zero.
Equation~\ref{eq:ssim_l} is qualitatively consistent with Weber's law.
Second, the standard deviation ($\sigma_{x_t}$ and $\sigma_{x_r}$) is employed as a round estimation of the signal contrast.
The contrast similarity function takes a similar form as luminance comparison
\begin{equation}\label{eq:ssim_c}
    c(\textbf{x}_t, \textbf{x}_r) = \frac{2\sigma_{x_t}\sigma_{x_r} + C_2}{\sigma_{x_t}^2 + \sigma_{x_r}^2 + C_2},
\end{equation}
where $C_2$ is another stabilization constant.
Similarly, the function qualitatively satisfies the contrast-masking feature of the HVS.
Third, the structure of distorted and reference images are defined as the normalized signals ($\textbf{x}_t-\mu_{x_t})/\sigma_{x_t}$ and ($\textbf{x}_r-\mu_{x_r})/\sigma_{x_r}$, respectively.
It should be noted that the formulation is in accordance with the initial definition that structural distortion is independent of nonstructural distortion.
The structure comparison function is defined as follows
\begin{equation}\label{eq:ssim_s}
    s(\textbf{x}_t, \textbf{x}_r) = \frac{\sigma_{x_tx_r} + C_3}{\sigma_{x_t}\sigma_{x_r} + C_3},
\end{equation}
where $C_3$ and $\sigma_{x_tx_r}$ are a stabilization constant and the correlation coefficient between $\textbf{x}_t$ and $\textbf{x}_r$, respectively.
Finally, the SSIM index is defined as the product of the three terms in Equation~\ref{eq:ssim_l},~\ref{eq:ssim_c}, and~\ref{eq:ssim_s}.
To simplify the expression, $C_3$ is set to $C_2 / 2$, resulting in
\begin{equation}\label{eq:ssim}
    d_{\mathrm{SSIM}}(\textbf{x}_t, \textbf{x}_r) = \frac{(2\mu_{x_t}\mu_{x_r} + C_1) (2\sigma_{x_tx_r} + C_2)}{(\mu_{x_t}^2 + \mu_{x_r}^2 + C_1)(\sigma_{x_t}^2 + \sigma_{x_r}^2 + C_2)}.
\end{equation}
The SSIM index is usually applied locally due to the spatially varying image statistical features and image distortions.
The overall quality of an image is, by default, computed as the average score across all local windows, though various spatial weighting strategies may be applied, many of which are shown to help improve the quality prediction accuracy~\cite{wang:06c,wang:10,zhang:16}.

The SSIM scores of the ``Barbara'' image set is shown in Figure~\ref{fig:example}, from which we can observe that the SSIM index correlate well with human quality perception.
Figure~\ref{fig:example}h shows the SSIM quality map for the noisy image, where brighter indicates better quality.
The noise over the region of the subject's face appears to be much stronger than that in the textured regions.
However, the MSE map is completely independent of the underlying image structures.
By contrast, the SSIM map gives perceptually consistent prediction.

Motivated by the success of SSIM, several variant models have been proposed by incorporating knowledge about visual psychophysics.
Most of them apply the SSIM index in the sub-band at different spatial locations~\cite{wang:10}, orientations~\cite{sampat:09,zhang:13}, and frequency content~\cite{wang:03,xue:13b,zhang:11} to simulate the characteristics of primary visual cortex.
Regardless of its simplicity and the empirical nature of the SSIM formulation, SSIM and its variations perform somewhat surprisingly well in various IQA tests.
For example, in a recently published and the most comprehensive IQA performance comparison so far, based on a collection of public domain IQA databases, almost all individual top-performing FR IQA methods were SSIM variations~\cite{athar:19}.

Another line of research explores alternative definitions of structure.
Indeed, the definition of structural/non-structural distortions is not unique.
For example, Wang~\textit{et~al.} extended the scope of non-structural distortions to non-linear luminance transformations and geometric image transformations~\cite{wang:05a}.
Recently, Ding~\textit{et~al.} defined texture resampling (\textit{e.g.}, replacing one patch of grass with another) as another instance of non-structural distortion~\cite{ding:20}. 

\subsection{Task-oriented Feature Learning Methods}
The structural similarity paradigm is conceptually appealing in the sense that it somehow by-passes the natural image complexity problem and the HVS complexity problem.
Indeed, these systems treat the HVS as a black box, and only the input-output relationship is of concern.
However, there is no simple unique answer on how to define structure and structural distortion in a perceptually meaningful manner.
Furthermore, there is no clear way to define and validate the optimality of the similarity measure $d(\textbf{x}_t, \textbf{x}_r; \bm{\theta})$.
To extend the structural similarity paradigm, other task-driven approaches have been introduced in the past decade, which differ from the structure similarity idea in two important ways.
First, the HVS are associated with more well-defined auxiliary tasks such as image recognition and semantic segmentation, as opposed to extracting structural information from the viewing field.
Second, the similarity measure is optimized using supervised machine learning methods.

Given some data in a multi-task setting, the task-driven methods estimate the prior distribution $p(\bm{\theta})$ by integrating out the task-specific parameters to form the marginal likelihood of the data.
Formally, grouping all of the data from each of the tasks as ${\bf \hat{X}}$ and again denoting by ${\bf \hat{x}}_{j1}, . . . , {\bf \hat{x}}_{j\hat{N}}$ a sample from task $\mathcal{T}_j$, the marginal likelihood
of the observed data is given by
\begin{equation}\label{eq:fr_task}
    p({\bf \hat{X}} | \bm{\theta}) = \prod_j \left(\int p({\bf \hat{x}}_{j1}, . . . , {\bf \hat{x}}_{j\hat{N}} | \bm{\psi}_j) p(\bm{\psi}_j | \bm{\theta}) d\bm{\psi}_j \right),
\end{equation}
where $\bm{\psi}_j$'s denote the task specific parameters.
Maximizing Equation~\ref{eq:fr_task} as a function of $\bm{\theta}$ gives a point estimate for $\bm{\theta}$, an instance of a method known as empirical Bayes~\cite{bernardo:09}.
Let $h({\bf x}_t; \bm{\theta})$ and $h({\bf x}_r; \bm{\theta})$ denote the feature representations of a pair of distorted image ${\bf x}_t$ and reference image ${\bf x}_r$ computed by the task-oriented function, the perceptual similarity index between the image pair is defined as
\begin{equation}\label{eq:task_dist}
    d_{\mathrm{Task}}({\bf x}_t, {\bf x}_r; \bm{\theta}) = d_{W} \big(h({\bf x}_t; \bm{\theta}), h({\bf x}_r; \bm{\theta})\big),
\end{equation}
where $d_W(\cdot, \cdot)$ is a certain distance measure in the feature domain, which may be either hand-crafted (\textit{e.g.}, the Euclidean distance~\cite{johnson:16,zhang:18} or multi-scale SSIM~\cite{gao:17}), or learnt from subjective rated images in a maximum a posterior manner~\cite{bosse:17}.
By leveraging the abundant training data in computer vision and the power of convolutional neural networks (CNN), these methods have demonstrated the potential to change the landscape of the field of IQA.


\subsection{Information Theoretic Paradigm}
The error visibility and the structural similarity paradigms have found nearly ubiquitous applications in the design of IQA systems, while both of them aim to derive a model for early sensory processing.
It turns out that there exists a distinct way to look at the IQA problem, \textit{i.e.} from the image formation point of view.
The information theoretic paradigm assumes that each reference image $\textbf{x}_r$ (usually its sub-images) is a sample from a very special probability distribution $p(\textbf{x}_r)$, \textit{i.e.}, the class of natural scenes.
Most real-world distortion processes disturb these statistics and make the image signal unnatural, suggesting that each distorted image $\textbf{x}_t$ comes from a distinct probability distribution $q(\textbf{x}_t)$.
As a result, the similarity between $\textbf{x}_t$ and $\textbf{x}_r$ can be measured by some information theoretic distance/divergence between these two probability distributions.

Although the use of information theoretic distances as perceptual similarity seems somewhat arbitrary, there exists a non-trivial connection between the two concepts.
Specifically, it has long been hypothesized that the HVS is adapted to optimally encode the visual signals~\cite{barlow:61,parraga:00}.
Because not all signals are equally likely, it is natural to assume that the perceptual systems are geared to best process those signals that occur most frequently.
Thus, the statistical properties of natural scene have a direct impact to the characteristics of the HVS.
Indeed, the statistical image modeling is shown to be the dual problem of the error visibility-based perceptual models~\cite{sheikh:05}.

To implement this idea, one has to specify the mathematical forms of natural image distribution $p(\textbf{x}_r; \bm{\theta}_1)$, distorted image distribution $q(\textbf{x}_t; \bm{\theta}_2)$, and the information theoretic distance measure $d_{\mathrm{INFO}}\big(p(\textbf{x}_r; \bm{\theta}_1), q(\textbf{x}_t; \bm{\theta}_2); \bm{\theta}_3\big)$, where we have represented our prior knowledge about the source image and the distortion process by $\bm{\theta} = \{\bm{\theta}_1, \bm{\theta}_2, \bm{\theta}_3\}$.
Furthermore, the problem of estimating $p(\textbf{x}_t; \bm{\theta}_1)$ and $q(\textbf{x}_r; \bm{\theta}_2)$ from a single sample is severely ill-posed.
To simplify the problem, it is often assumed that image statistics are locally homogeneous and the patches within an image are independent and identically sampled from the corresponding distribution.
The probability distributions are then estimated from a stack of sub-images within the pair of distorted and reference images.
All information theoretic IQA methods can be explained by the framework, although they differ in detail.

As an initial attempt in this paradigm, the Information Fidelity Criterion~\cite{sheikh:05} models the natural image distribution $p(\textbf{x}_r; \bm{\theta}_1)$ as a Gaussian Scale Mixture~\cite{wainwright:00}.
To derive the model for the distorted image distribution $q(\textbf{x}_t; \bm{\theta}_2)$, the method assumes the distortion process to consist a simple signal attenuation and additive Gaussian noise.
Finally, the perceptual quality is measured by the mutual information~\cite{cover:91} between $p(\textbf{x}_r; \bm{\theta}_1)$ and $q(\textbf{x}_t; \bm{\theta}_2)$.
As a close variant of the Information Fidelity Criterion, Visual Information Fidelity (VIF) approaches the HVS as a ``distortion channel'', which introduces stationary, zero mean, additive white Gaussian noise to the images in the wavelet domain~\cite{sheikh:06b}.
Other extended version have adopted other statistical models as the image density model~\cite{chang:13,wang:05b,wang:06b}, estimated the image distributions in other transform domains~\cite{wang:06b}, and employed other probabilistic distance measure as the perceptual similarity measure~\cite{rehman:12,soundararajan:11,wang:06b}.

Figure~\ref{fig:example} shows the prediction results of VIF on a set of altered ``Barbara'' images.
In comparison with the reference image, the contrast enhanced image has a better visual quality despite the fact that the `distortion' (in terms of a perceivable difference with the reference image) is clearly visible.
A VIF value larger than unity captures the improvement in visual quality.
In contrast, the noisy image, the blurred image, and the JPEG compressed image have clearly visible distortions and poorer visual quality, which is captured by a low VIF measure for all three images.
The quality map predicted by VIF in Figure~\ref{fig:example}l is also consistent with human perception.

Despite the demonstrated success, the information theoretic paradigm suffers from two notable limitations.
First, the independent and identically distributed assumption barely holds in practice, since neighboring spatial locations are strongly correlated in intensity~\cite{simoncelli:01}.
Second, many methods makes explicit/implicit assumptions about the distortion process in order to determine the distorted image distribution.
However, given a distorted image $\textbf{x}_t$ and a reference image $\textbf{x}_r$, the image quality $y$ is independent of the distortion process.
The unnecessary assumption about the distortion process introduces inductive bias to the IQA models, resulting in less competitive generalization capability.

\subsection{Fusion-based Methods}
All of the paradigms above are well-motivated, and have achieved great success in predicting subjective quality perception~\cite{sheikh:06a}.
However, it has been demonstrated that the performance of these methods fluctuate across different distortions~\cite{athar:19}.
Given the diversity of knowledge sources, a natural question is how to make use of different sources of knowledge in one IQA model.
To this regard, fusion-based IQA methods are developed to build a ``super-evaluator'' that exploits the diversity and complementarity of the existing methods for improved quality prediction performance.

Given $l$ point estimate of model configurations $\{\bm{\theta}_k\}_{k=1}^l$, most fusion-based methods can be explained by a ``mixture of experts'' model.
The approach assumes the posterior quality distribution have a hierarchical form
\begin{equation}\label{eq:fusion}
    p(y|\textbf{x}_t,\textbf{x}_r,\bm{\theta}) = \sum_{k=1}^l p(y|\textbf{x}_t,\textbf{x}_r,z=k,\{\bm{\theta}_k\}_{k=1}^l) p(z=k|\textbf{x}_t,\textbf{x}_r,\{\bm{\theta}_k\}_{k=1}^l),
\end{equation}
where each image has an unknown class $z$, $p(y|\textbf{x}_t,\textbf{x}_r,z=k,\{\bm{\theta}_k\}_{k=1}^l)$ is the $k$-th base IQA model, and $p(z=k|\textbf{x}_t,\textbf{x}_r,\{\bm{\theta}_k\}_{k=1}^l)$ weights the predictions of each ``expert'' in an ensemble.
Due to the lack of training data, early researches assume class conditional distribution to be independent of the input image pair.
The form of latent variable distribution $p(z=k|\{\bm{\theta}_k\}_{k=1}^l)$ can be determined empirically~\cite{ye:14} or learnt from data~\cite{liu:12,ma:19}.
There have also been attempts in getting rid of the independence assumption, which unfortunately achieved less impressive results~\cite{athar:19}.


\subsection{Discussion}
\noindent\textbf{The Relationship between Image Fidelity and Image Quality}:
The equivalence between image quality and image fidelity relies on a few critical assumptions.
First, it is assumed that the reference image is of perfect quality.
If the assumption is violated, an image can sometimes be ``enhanced'' by a distortion.
Observers may detect the difference between an original and its distorted version and prefer the distorted version over the original.
Second, it is often assumed that there is at least a proportional relationship between the visibility of the distortion and the difference in perceived quality of the image~\cite{silverstein:96}.
The assumption may hold for high fidelity, but often fails at low fidelity levels, for example, an image with distinct content could still have a perfect image quality.
Furthermore, this assumption does not always hold in practice as certain distortion type may be clearly visible but not so objectionable.

\noindent\textbf{The Quality Definition Problem}: Perhaps an even more fundamental problem with the FR IQA models is the definition of image quality.
The definition of image quality depends on the definition of pristine image, which usually refers to the image with perfect quality.
However, image quality has to be defined in the first place in order for the definition to take effect.
Apparently, this has run into a circular definition problem.

\section{No-Reference Image Quality Assessment}
No-reference (NR) IQA models aim to directly evaluate the quality of an image without referring to an ``original'' high-quality image.
The task is in general extremely challenging for artificial vision systems.
Yet, amazingly, this is quite an easy task for human observers.
Human observers can easily identify high-quality images versus low-quality images and detect distortions in an image.
Furthermore, humans tend to agree with each other to a high extent.
These evidences suggest that it is possible to develop a machine vision system to perform NR IQA, though discovering the mechanisms underlying human perceptual IQA is highly challenging.

\begin{figure}[t!]
\includegraphics[width=\textwidth]{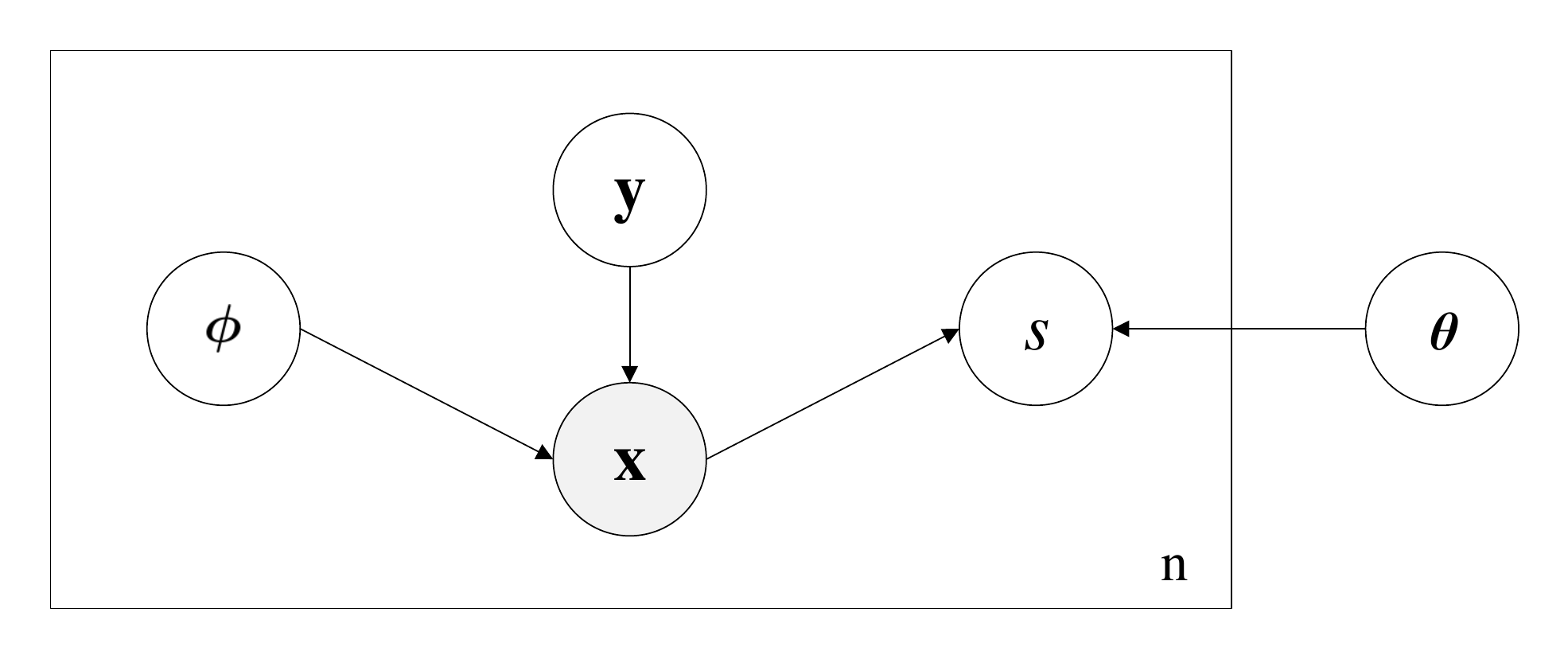}
\caption{Graphical model representation of NR IQA models.}
\label{fig:nriqa}
\end{figure}

The NR IQA problem can also be explained by Equation~\ref{eq:prediction} $p(y|{\bf x}, \mathcal{D}) = \int p(y|{\bf x}, {\bm{\theta}}) p(\bm{\theta}|\mathcal{D}) d\bm{\theta}$, where each observation ${\bf x}$ consists of only a test image ${\bf x}_t$.
The probabilistic graphical model of NR IQA models is shown in Figure~\ref{fig:nriqa}, where we observe two differences from the FR IQA models.
First, the original image ${\bf x}_r$ is not observable.
Second, the quality score $y$ is assumed to be independent of the reference image ${\bf x}_r$ conditioned on the test image ${\bf x}_t$.
Over the past decade, a great number of NR IQA models have been developed, which may be broadly classified into three categories.

\subsection{Empirical Statistical Modeling Approach}
It has long been conjectured, with abundant supporting evidence, that the role of early biological sensory systems is to remove redundancies in the sensory input, resulting in a set of neural responses that are statistically independent, known as the ``efficient coding'' principle~\cite{barlow:61}.
Assuming that the visual systems have evolved to become optimal and more ``comfortable'' working with familiar input signals, it follows that an image appearing more frequently in the natural world, or in other words more ``natural'', would have better visual quality.
To fully specify the hypothesis, one needs also to state which environment shapes the system.
Quantitatively, this means specification of a probability distribution over the space of input signals.
Following this philosophy, significant efforts have been devoted to determine the prior parameter distribution $p({\bm{\theta}})$ by estimating the probability density function of test images $p({\bf x}_t|\bm{\theta})$ (and natural images $p({\bf x}_r|\bm{\theta})$).

The density estimation problem is very challenging due to the fundamental conflict between the enormous size of the image space and the limited number of images available for observation.
There have been two techniques to alleviate the problem, which are summarized as follows:
\begin{itemize}
    \item Dimension Reduction with Hierarchical Model: One method that has been demonstrated to be useful is dimension reduction.
    The idea is to map the entire image space onto a space of much lower dimensionality by exploiting knowledge of the statistical distribution of ``typical'' images in the image space.
    Since natural images have been found to exhibit strong statistical regularities~\cite{simoncelli:01}, it is possible that the cluster of typical natural images may be represented by a low-dimensional manifold, thus reducing the number of sample images that might be needed in the subjective experiments.
    The dimension reduction approach corresponds to a specific family of image density models
    \begin{equation}\label{eq:dim_reduction}
    p_\textbf{x}(\textbf{x}_t; \bm{\theta}) = \int p(\textbf{x}_t|\textbf{z}; \bm{\theta}_1) p(\textbf{z}; \bm{\theta}_2) d\textbf{z},
    \end{equation}
    where $\textbf{z}$ is a low dimensional latent variable, and $\bm{\theta} = \{\bm{\theta}_1, \bm{\theta}_2\}$.
    The probability distribution of pristine images $\textbf{x}_r$ can be modeled either jointly with distorted images $\textbf{x}_t$~\cite{mittal:12a,saad:12}, or independently as a separate model~\cite{mittal:12b,moorthy:11,wu:15b}.
    For example, the conditional probability distribution $p(\textbf{x}_t|\textbf{z}; \bm{\theta}_1)$ is often modeled by an Asymmetric Generalized Gaussian distribution~\cite{lasmar:09} in a localized linear transform domain, where spatially distant pixels are assumed to be uncorrelated for simplicity.
    The reduced sample space in $\textbf{z}$ makes it possible to learn the probability density $p(\textbf{z}; \bm{\theta}_2)$ from data.
    To avoid under-fitting, most existing algorithms estimate $p(\textbf{z}; \bm{\theta}_2)$ in a non-parametric manner, which makes few assumptions about the form of the distribution.
    Alternative methods apply the dimension reduction $p(\textbf{x}_t|\textbf{z}; \bm{\theta}_1)$ on medium-sized image patches, and learn a parametric $p(\textbf{z}; \bm{\theta}_2)$ model in order to obtain a generative model with explicit mathematical expression~\cite{hou:14,mittal:12b,wu:15a,zhang:15a}.
    For example, a representative method called NIQE~\cite{mittal:12b} use the Asymmetric Generalized Gaussian distribution to fit $p(\textbf{x}_t|\textbf{z}; \bm{\theta}_1)$ by $96 \times 96$ image patches, and assume that the latent variable $\textbf{z}$ follows a multi-variate Gaussian distribution.
    \item Patch-based Density Estimation: It should be noted that the aforementioned natural image statistic models remain overly simplistic, in the sense that they yield insufficiently adequate descriptions of the probability distribution of natural images in the space of all possible images.
    To overcome the limitation, an alternative method directly learns the probability density function of low-dimensional sub-images by assuming that the image patches are independent and identical samples of $p(\textbf{x}_t|\bm{\theta})$ (or $p(\textbf{x}_r|\bm{\theta})$ if the patches come from a pristine image).
    The research in IQA is constantly searching for the optimal form of the probability distribution.
    A pioneering method following this approach named CORNIA~\cite{ye:12} jointly models the probability distribution of both natural images and distorted images by a Gaussian Mixture Model.
    Despite its simplicity, CORNIA remains as one of the most competitive NR IQA models~\cite{athar:19}.
    Follow-up works have demonstrated that marginal improvements can be attained by using more powerful probability mixture models~\cite{xu:16}.
\end{itemize}

Despite the proven efficiency, both approaches make over-simplified empirical assumptions about the image density, which inevitably reduces their accuracy.
Over the past five years, we have witnessed an exponential growth in research activity into the advanced training of purely data-driven models.
Thanks to the availability of significantly larger data sets and the dedicated hardware unit that can efficiently process large volume of data, it becomes possible to learn a high dimensional image density model with exact log-likelihood computation, exact and efficient sampling, exact and efficient inference of latent variables, and an interpretable latent space~\cite{kingma:18}.
These models have demonstrated a significant improvement in log-likelihood on standard benchmarks over the traditional approaches without relying on excessive assumptions.
It remains to be seen how much these models can improve the performance of the current NR IQA algorithms.

\subsection{Fidelity Model Distillation Approach}
Inspired by the remarkable achievement of FR IQA techniques over the past decade, several studies proposed to directly learn the prior distribution from FR IQA models in hope that the NR models could inherit the prior knowledge from them.
There exist two sub-categories in the fidelity model distillation method, which differ in their way to make use of FR IQA models.

\begin{itemize}
    \item Learning from Synthetic Quality Labels: The first approach directly adopts the quality prediction of FR IQA models as the ground-truth label and learns the prior distribution in a supervised learning fashion.
    Given a dataset of $n$ pristine images $\textbf{X}_r = (\textbf{x}_{r,1}, ..., \textbf{x}_{r,n})$, a distortion simulator $g(\cdot; {\bm \phi})$, and a FR IQA model $d(\textbf{x}_t, \textbf{x}_r)$, the fidelity model distillation approach firstly generates a set of synthetically distorted images $\textbf{X}_t = (\textbf{x}_{t,1}, ..., \textbf{x}_{t,n})$, where $\textbf{x}_{t,i} = g(\textbf{x}_{r,i}; {\bm \phi})$.
    For each pair of distorted and reference images $(\textbf{x}_{t,i}, \textbf{x}_{r,i})$, a synthetic quality score $\hat{y}_i = d(\textbf{x}_{t,i}, \textbf{x}_{r,i})$ is then derived from the FR IQA measure, which can be denoted collectively as $\hat\textbf{y}$.
    Assuming the generated data are independent and identically distributed, the prior model parameter $\bm{\theta}$ is set to the value that maximizes the likelihood function $p(\textbf{X}_t, \hat\textbf{y}|\bm{\theta})$.
    Various instantiations of the idea have been developed based on different FR IQA models.
    Many algorithms are built upon standalone FR IQA models for conceptual simplicity~\cite{kim:16,kim:18,pan:18}.
    To take advantage of all three types of knowledge sources, state-of-the-art models of this kind employ fusion-based FR IQA models as the quality annotator~\cite{wang:19,ye:14}.
    These models yield high correlation with human opinion scores on the standard distorted images whose distortion process can be faithfully simulated.
    \item Learning to Rank: During the data preparation stage, the distortion simulator typically generates multiple distorted images for each reference image to cover the diversity of distortion processes, suggesting that the training data are not independent and identically distributed.
    To mitigate the problem, other fidelity model distillation-based models learn from the relations among the training images.
    Specifically, for each pair of images $(\textbf{x}_{t,i}, \textbf{x}_{t,j})$ in the training set, let $r_{ij}=1$ if $\hat{y}_i > \hat{y}_j$ and $r_{ij}=0$ otherwise.
    Assuming the variability of quality across images is uncorrelated, the reliability of the IQA annotators do not depend on the input image, and the image pairs in the dataset are independent and identically distributed~\cite{ma:19},
    one can then obtain the prior parameter distribution $\bm{\theta}$ by maximizing the likelihood function 
    \begin{equation}\label{eq:l2r}
    p(\{\textbf{x}_{t,i}, \textbf{x}_{t,j}, r_{ij}\}| \bm{\theta}) = \prod_{\langle i,j \rangle} p(r_{ij}|\textbf{x}_{t,i}, \textbf{x}_{t,j},\bm{\theta}).
    \end{equation}
    To fully specify the optimization problem, one also need to make assumptions about the mathematical form of $p(r_{ij}|\textbf{x}_{t,i}, \textbf{x}_{t,j},\bm{\theta})$.
    Early attempts of this approach models the conditional probability with some standard functions (\textit{e.g.}, step function, standard Normal cumulative distribution function)~\cite{gao:15}, while state-of-the-art algorithms employ hierarchical probabilistic models for better model capacity~\cite{ma:17} and interpretability~\cite{ma:19}.
\end{itemize}

In general, the fidelity model distillation-based NR IQA models have to face three major challenges.
First, the robustness of this approach heavily relies on the diversity and quality of the synthetic distortion generator,
both of which are often questionable in practice.
Specifically, only a dozen of distortion types may be simulated, which may be inadequate at representing the diversity of distortions.
As a result, this type of models does not generalize well to out-of-distribution distortion types~\cite{athar:19}.
Second, their performance is upper-bounded by that of FR IQA models, which may be inaccurate across distortion levels~\cite{ma:20} and distortion types~\cite{ponomarenko:15}.
Third, even if the target FR IQA model performs perfectly on the synthetic distorted image dataset, the approach may suffer from excessive label noise originated from the natural discrepancy between perceptual fidelity and image quality.
In particular, a distorted image could correspond to several plausible pristine counterparts, resulting in drastically different perceptual similarity measurements.
Without access to the actual original images, the learner may be confused by the diverse quality annotations during the training stage.

\subsection{Transfer Learning Approach}
This approach is essentially the NR counterpart of the task-oriented feature learning methods for FR IQA.
The basic assumption is that the HVS parameter configuration optimized for one visual task may also perform well on a relevant task.
Methods of this kind maximize Equation~\ref{eq:fr_task} on various visual tasks via maximum likelihood method to obtain a prior estimate for $p(\bm{\theta})$, upon which the posterior distribution is derived.
The instantiations of the approach differs in their domain of supplementary tasks.

Motivated by the prevalence of deep learning, most transfer learning-based IQA methods approximate the marginal likelihood of the observed data in the auxiliary task domain with a CNN.
When developing the IQA models, researchers typically freeze the convolutional layers optimized for an auxiliary task (which are not retrained), and only retrain the fully connected layers that implement IQA circuits at the top to associate visual representations derived from the convolutional layers with quality annotations.
Alternatively, the convolutional layers may be initialized with the auxiliary task-optimized parameters, and are fine-tuned by subject-rated images via a few gradient descent steps.
The learning method is equivalent to an empirical Bayes procedure to maximize the marginal likelihood that uses a point estimate for $\bm{\theta}$ computed by one or a few steps of gradient descent.
However, this point estimate is not necessarily the global mode of a posterior due to the non-linearity of the CNN.
We can instead understand the point estimate given by truncated gradient descent as the value of the mode of an implicit posterior over $\bm{\theta}$ resulting from an empirical loss interpreted as a negative log-likelihood, and regularization penalties and the early stopping procedure jointly acting as priors~\cite{grant:18}.
It is worth mentioning that the CNN architecture itself has been imposed as the prior knowledge about the connectivity of neurons in primary visual cortex.

The earliest transfer learning-based NR IQA models employ image recognition~\cite{bosse:17,bianco:18} as the auxiliary task, where abundant subject annotations exist~\cite{russakovsky:15}.
Somewhat surprisingly, the pre-trained network already exhibits moderate correlation with subjective quality annotations, suggesting that the task-oriented visual representations are to some degree already quality-aware~\cite{kim:17b}.
With minimal fine-tuning, the method achieves much better performance.
Another model are optimized in a similar fashion with the pre-training task being image restoration~\cite{lin:18}.
The performance and efficiency of these approaches depend highly on the generalizability and relevance of the tasks used for pre-training.
To enhance the relevance of the auxiliary task to IQA, a few recent algorithms have the quality prediction sub-task regularized by distortion identification~\cite{kang:15,ma:18}.
However, the method is not easily extended for authentically distorted images because there is no well-defined categorization of real-world image distortions.
Furthermore, it remains unclear if the HVS performs distortion identification as a explicit visual task.
The search for optimal auxiliary tasks in the context of IQA is a subject of ongoing research.


\subsection{Discussion}
\noindent\textbf{The Knowledge about Distortion Process}: The knowledge about distortion process has played an important role in many IQA models, especially in the case of application-specific IQA where efficient algorithms may be developed by assessing the severeness of certain distortions.
In the case of general-purpose IQA, however, the use of such knowledge may not be preferable for the following reasons.
First, the development of universal distortion model is extremely challenging, because of the constantly evolving distortion process distribution.
Indeed, the distortions that can occur are infinitely variable and one cannot predict whether or not a hitherto-unknown distortion type will emerge tomorrow.
To account for all possible distortion types, one may have to assume a uniform distribution of the distortion process, which is equivalent to not using any knowledge about image distortions~\cite{wang:06a}.
Second, a na\"ive subject can consistently assess image quality without access to the underlying distortion process, suggesting that the visual systems are capable of judging image quality independent of the knowledge about distortion.
By contrast, existing NR IQA methods make use of the knowledge about image distortions in some way (\textit{e.g.}, by assuming the probability density function of distorted images, predicting the distortion type as an auxiliary visual task, or using distortion simulator to generate training data).

\noindent\textbf{The Data Challenge}: The success of IQA models strongly depends on the quantity, quality, representativeness, and consistency of training data, all of which are extremely limited in practice.
First, the quantity of subject-rated images is bounded by the small capacity for subjective measurements.
A typical ``large-scale'' subjective test allows for a maximum of several hundreds or a few thousands of test images to be rated.
Given the enormous space of digital images, a few thousands of subject-rated samples are deemed to be extremely sparsely distributed in the space.
Second, the quality of subject ratings is inherently lower than the labels in other visual tasks such as image categorization and segmentation due to the stochastic nature of image quality.
More importantly, the quality of subject ratings gradually degrades as the number of test samples in a subjective experiment increases, where the fatigue effect comes into play.
Third, the subject-rated images in the existing IQA databases may not be representative of the real-world distorted images, whose distortion process cannot be faithfully reproduced.
Fourth, the consistency of subjective image quality among IQA databases is only moderate due to drastically different experimental conditions.
Strictly speaking, the quality ratings of an image $\textbf{x}_t$ collected from a subjective experiment are essentially samples from a context conditional quality distribution $p(y|\textbf{x}_t, \textbf{t})$, where $\textbf{t}$ encodes the information about experiment environment, instruction, training process, presentation order, and experiment protocol.
As a result, the subjective quality ratings obtained from different experiments cannot be simply aggregated into a larger IQA dataset $p(y|\textbf{x}_t)$.
These data challenges constantly arise in IQA research and will remain a challenging issue in the future.

\noindent\textbf{The Fair Comparison Challenge}: Given the diversity of design philosophies, it becomes very challenging to fairly compare two competing hypotheses.
Specifically, existing IQA algorithms are often trained on different datasets, equipped with different model capacity, and optimized by different learning algorithms.
It remains unclear whether the performance gain comes from a more representative dataset, a more powerful model, an advanced machine learning technique, or the superiority of the proposed hypothesis.
To ascertain the improvement, we expect more controlled experiments in the future.

\noindent\textbf{The Cognitive Interaction Problem}: It is widely known that cognitive understanding and interactive visual processing (\textit{e.g.}, eye movements) influence the perceived quality of images.
For instance, the subjective quality rating of an image is shown to be a function of the experiment instruction~\cite{sheikh:06a}.
The preference of image content, prior information about image composition, or attention and fixation~\cite{engelke:11,zhang:16} may also affect the evaluation of the image quality.
The incorporation of cognitive process in the IQA is a subject of ongoing research~\cite{liu:11,zhang:17}.


\begin{figure}[t!]
\includegraphics[width=0.7\paperwidth]{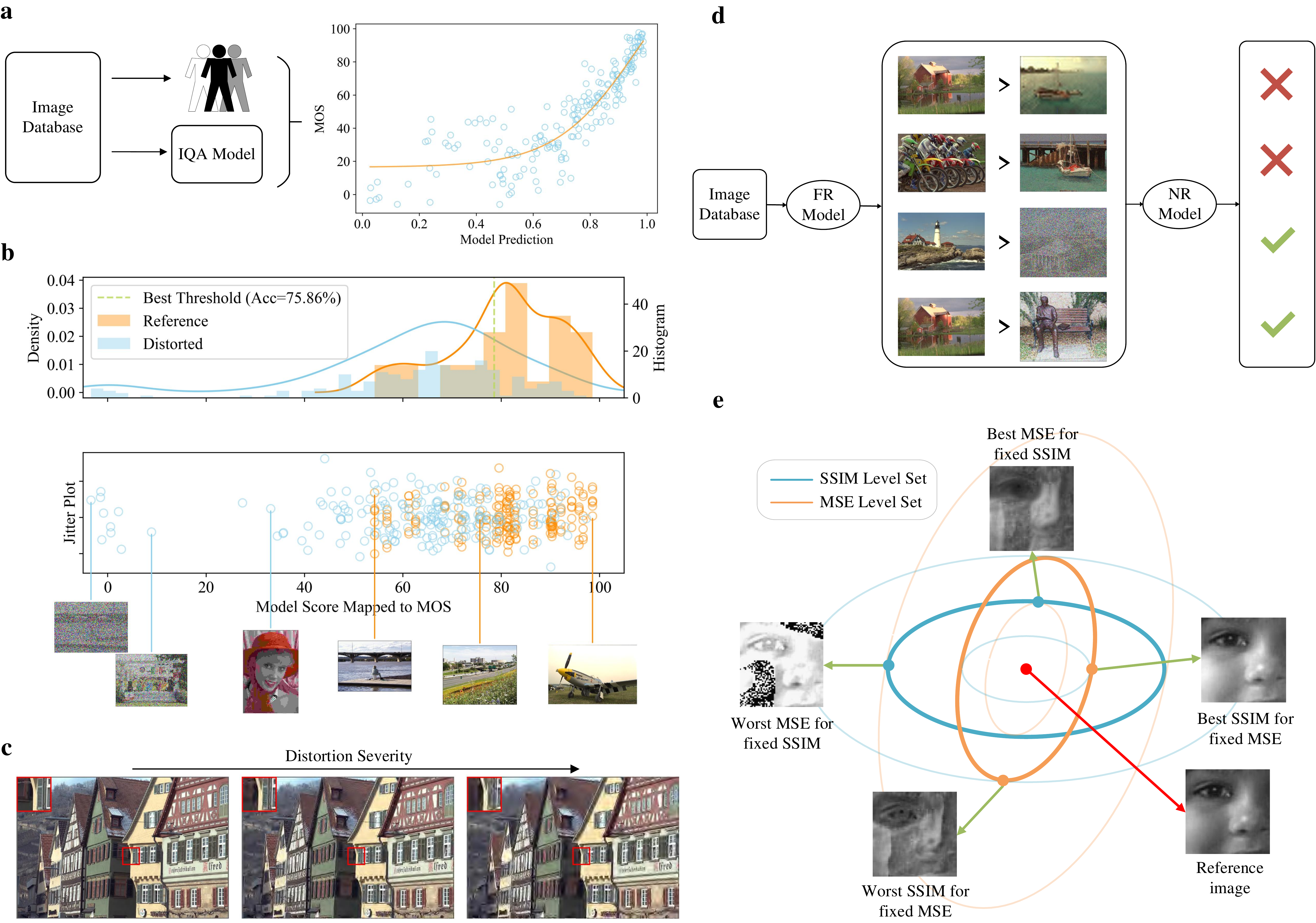}
\caption{Existing evaluation procedures for objective IQA models. (a) Direct correlation with subjective evaluation: The objective model predictions are directly compared to subjective annotations on a database of images. (b) D-Test: NR IQA models are evaluated based on their capability to separate distorted images from pristine ones. (c) L-Test: NR IQA models are tested to identify the severity of synthetic distortions. (d) P-Test: NR IQA models are evaluated by their ability to identify discriminable image pairs. (e) MAD stimulus synthesis in the image space.}
\label{fig:validation}
\end{figure}

\section{Evaluation Methodology}
With a significant number of IQA models proposed recently, how to fairly compare their performance becomes a challenge.
The existing evaluation methodologies are summarized in Figure~\ref{fig:validation}, and discussed in detail below:

\noindent\textbf{Direct Correlation with Subjective Evaluation}: Because the HVS is the ultimate receiver in most applications, subjective evaluation is a straightforward and reliable approach to evaluate image quality.
The method constitutes three steps as illustrated in Figure~\ref{fig:validation}a.
In the first stage, a number of representative images are selected from the image space.
Early researches collect a few dozens of pristine images and distort the source images with distortion simulators that create distorted images of a few pre-set distortion types and quality levels~\cite{larson:10,ponomarenko:15,ponomarenko:09,sheikh:06a}.
However, the real-world image distortion may deviate significantly from such simulated images.
To this regard, recent studies create datasets of real-world Internet images, which are contaminated by authentic distortions~\cite{ghadiyaram:15,hosu:20}.
In the second stage, the selected images are evaluated by a number of subjects.
Each subject gives a quality score to each selected image, and the overall subjective quality of the image is typically represented by its mean opinion score (MOS)~\cite{sheikh:06a}.
Alternatively, the subjective experiment may be setup in a double stimulus setting, where subjects are provided with two images and are asked to select the one with better quality.
The preference data can be aggregated into a global ranking using rank aggregation tools such as maximum likelihood for multiple options~\cite{ponomarenko:15,ponomarenko:09}.
In the final stage, the performance of the objective models is evaluated by comparison with subjective scores.
Typical evaluation criteria include (1) Pearson linear correlation coefficient after a non-linear monotonic mapping between objective and subjective scores: a parametric measure of prediction accuracy;
(2) Spearman rank-order correlation coefficient: a non-parametric measure of prediction monotonicity;
and (3) Kendall rank-order correlation coefficient: another non-parametric measure of prediction monotonicity.
A major problem with this evaluation methodology is the conflict between the enormous size of the image space and the limited capacity for subjective experiment.
Subjective testing is expensive and time-consuming.
The largest IQA dataset contains only 10,000 subject-rated images, which are deemed to be sparse samples of the image space.

\noindent\textbf{Rational Test}: NR IQA models can also be evaluated in a more economic way without conducting subjective experiment.
Existing objective evaluation criteria rely on an image database consisting of pristine images and the synthetic distorted images derived from them.
\begin{itemize}
    \item Pristine/Distorted Image Discriminability Test (D-Test)~\cite{ma:16}: The procedure of D-Test is shown in Figure~\ref{fig:validation}b.
    Considering the pristine and distorted images as two distinct classes in a meaningful perceptual space, the D-Test aims to test how well an IQA model is able to separate the two classes.
    For each test IQA model, the procedure seeks a threshold value optimized to yield the maximum correct classification rate.
    A good NR IQA model should accurately distinguish the pristine images from the distorted ones.
    \item Listwise Ranking Consistency Test (L-Test)~\cite{winkler:12}: The goal is to evaluate the robustness of IQA models when rating images of the same content and with the same distortion type but different distortion levels.
    A good IQA model should rank these images in the same order.
    An illustrative example is given in Figure~\ref{fig:validation}c, where different models may or may not produce the same quality rankings in consistency with the image distortion levels.
    The method assumes that the quality of an image degrades monotonically with the increase of the distortion level for any distortion type, which may not generalize to all distortion processes (\textit{e.g.}, rotation, contrast change, etc.).
    \item Pairwise Preference Consistency Test (P-Test)~\cite{ma:16}: The evaluation method relies on FR IQA models to select image pairs whose quality is clearly discriminable.
    In contrast to L-Test, this evaluation criteria enables the comparison of IQA models in their cross-content capability.
    In practice, an image pair is considered to be discriminable in quality if the difference in FR IQA predictions is larger than a certain threshold.
    The flowchart of P-Test is illustrated in Figure~\ref{fig:validation}d.
    A good NR IQA model should consistently predict preferences concordant with the discriminable image pairs.
    The underlying assumption is that the target FR IQA generalize well to the synthetic distortions.
\end{itemize}
The dependence of these rational tests on distortion simulators limits their effectiveness as a strong benchmark, as a NR IQA model succeeding the sanity check may fail on authentically distorted images.
Nevertheless, the objective evaluation methods provide an economic complement to the standard subjective evaluation, which have demonstrated to be especially useful in training machine learning-based NR IQA models.

\noindent\textbf{Analysis by Synthesis}: Given the enormous size of the image space, the limited capacity for subjective experiment, and the constantly evolving distortion processes, it seems hopeless to verify IQA models in a comprehensive manner.
By contrast, to fail a model can be maximally efficient, for which theoretically only one counterexample is sufficient.
Therefore, to accelerate the model comparison process, a complementary proposal is to falsify rather than validate the models.
The method dubbed MAximum Differentiation (MAD) competition is illustrated in Figure~\ref{fig:validation}e using MSE and SSIM as examples of competing models.
Given two IQA models, MAD competition searches for a pair of images that maximize/minimize the quality in terms of one model (termed the attacker model) while holding the other (termed the defender model) fixed.
The problem can be solved by advanced optimization algorithms~\cite{berardino:17,wang:04c,wang:08}, or exhaustive search in a large pool of pre-selected images~\cite{ma:20}.
Following the stimuli synthesis, a two alternative forced choice subjective experiment (or its variant) is carried out to disprove the defender model.
This procedure is then repeated, but with the attacker/defender roles of the two models reversed.
A defender model that better survives attacks from other models in such a MAD~\cite{wang:08} or group MAD~\cite{ma:20} competitions, or an attacker model that better attacks/fails other models in such competitions, is considered a better model.

\section{Conclusion and Open Problems}
We have presented a Bayesian view to the visual image quantification problem.
We have demonstrated that existing IQA methods can be explained by a common Bayesian framework with concrete mathematical formulation.
To facilitate the understanding and comparison of these approaches, we have made the underlying assumptions explicit.
Provided the ill-posed nature of IQA problem, it is essential to incorporate prior knowledge in the design of computational visual models.
Depending on the availability of the reference image, two types of probabilistic graphical model can be derived, which define image quality in different ways.
Both approaches aim to discover the configuration of the HVS represented by the prior distribution $p(\bm{\theta})$.
Despite the variations in design principles and the great diversity of modeling techniques, all existing methods make use of one or more of three types of prior knowledge: knowledge about the HVS; knowledge about high-quality images; and knowledge about image distortions.

Remarkable progress has been made in the past decades in the field of IQA, evidenced by a number of state-of-the-art IQA models achieving high correlations with subjective quality opinions on images when tested using publicly available image quality databases.
Nevertheless, this does not necessarily mean that IQA research has reached a level of maturity, especially when facing real-world challenges~\cite{chandler:13,wang:16a}.
First, existing IQA models often suffer from generalization problem.
It has been observed that the performance of IQA models trained on one database reduces significantly on other benchmark datasets, largely due to the distribution mismatch in the visual content and the distortion process across datasets.
The lack of generalized, reliable, and easy-to-use model validation procedure also hinders the development of truly successful IQA systems.
Second, most existing IQA models do not exhibit desirable mathematical properties, making it difficult to derive reliable perceptually motivated optimization approaches in image processing, computer vision, and computer graphics applications.
Only limited effort has been made on understanding the mathematical properties of IQA measures~\cite{brunet:11,brunet:12,richter:11}.
Third, it is highly desirable to reduce the complexity of IQA algorithms, especially for time-sensitive applications such as live broadcasting and video conferencing.
Many existing models are far from meeting this challenge.

It is worth noting that the IQA tasks discussed so far have been constrained to an ideal narrow scope that allows for a focused, in-depth discussion.
In practice, there is an enormous demand of IQA algorithms and systems, many of which involve novel domain-specific challenges.
The application scope includes, but is not limited to, computer graphics~\cite{lavoue:15}, video compression~\cite{zeng:14}, video streaming~\cite{duanmu:16}, camera process~\cite{fang:20}, printing~\cite{kite:00}, visual displays~\cite{rehman:15}, stereo vision~\cite{lambooij:11}, reduced-reference quality assessment~\cite{wang:11}, degraded-reference quality assessment~\cite{athar:17}, multi-exposure fusion~\cite{ma:15a}, dynamic range compression~\cite{yeganeh:13}, texture analysis~\cite{zujovic:13}, spatial interpolation~\cite{yeganeh:15}, video frame-rate conversion~\cite{nasiri:17}, color image reproduction~\cite{zhang:97}, color-to-gray conversion~\cite{ma:15b}, depth quality~\cite{wang:16b}, visual discomfort~\cite{lambooij:09}, image aesthetics~\cite{deng:17}, new media types and environment (virtual reality and augmented reality)~\cite{kim:19}, screen content~\cite{min:17}, point cloud~\cite{su:19}, and 360-degree omnidirectional content~\cite{xu:20}, among many others.
Most of these works are in preliminary stages, and there is a large space to be explored in the future.

\section{Summary Points}
\begin{enumerate}
\item Objective image quality assessment (IQA) can be formulated as a Bayesian inference problem, where the key is to obtain the configuration of the human visual system (HVS) encoded by a prior parameter distribution.
\item In general, three types of knowledge may be used in the design of image quality assessment methods: knowledge about the HVS; knowledge about high-quality images; and knowledge about image distortions.
\item Perceptual fidelity is closely related to image quality under certain conditions.
      Based on this observation, a variety of full-reference IQA models are developed, including the error visibility paradigm, the structural similarity paradigm, the information theoretic paradigm, task-oriented feature learning methods, and fusion-based methods.
\item No-reference IQA models can predict the visual quality of an image without access to its pristine counterpart.
      Existing methods can be categorized into the empirical statistical modeling approach, the fidelity model distillation approach, and the transfer learning approach.
\item There has been a recent trend in the design principles of IQA methods from knowledge-driven toward data-driven approaches, evident by the dominance of objective prior learnt by Empirical Bayes method over the subjective prior designed by IQA researchers.
\item The generalizability of IQA models, especially data-driven models, strongly depends on the quantity, quality, representativeness, and consistency of training data, which are scarce in practice.
      Creative methods are desired to mitigate such data challenges, and to overcome the limited capability of evaluation procedures.
\end{enumerate}

\subsubsection*{Acknowledgments}
This work is supported in part by Natural Sciences and Engineering Research Council (NSERC) of Canada under the Discovery Grant, Canada Research Chair program, and Alexander Graham Bell Canada Graduate Scholarship program.

The manuscript has been accepted by Annual Review of Vision Science.

Figure~\ref{fig:friqa}, Figure~\ref{fig:error_visibility}, and Figure~\ref{fig:nriqa} are absent in the accepted manuscript for conciseness.








\end{document}